\documentclass[review]{elsarticle}

\usepackage{lineno,hyperref}
\modulolinenumbers[5]

\usepackage{color}
\usepackage[normalem]{ulem}
\usepackage{diagbox}
\usepackage{makecell}

\journal{arXiv}

\begin{document}

\begin{frontmatter}

\title{New Empirical Evidence on Disjunction
Effect and Cultural Dependence}

\author{Indranil Mukhopadhyay\fnref{email1}}
\address{Human Genomics Unit, Indian Statistical Institute, Calcutta-700035.}
\fntext[email1]{ Email: indranilm100@gmail.com.}

\author{Nithin Nagaraj\fnref{email2a}}
\address{Consciousness Studies Programme, National Institute of Advanced Studies, IISc. Campus, Bengaluru 560012.}
\fntext[email2a]{Email: nithin@nias.iisc.ernet.in.}
\author{Sisir Roy\fnref{email2b}}
\address{Conciousness Studies Programme, National Institute of Advanced Studies,IISC Campus, Bengaluru 560012.}
\fntext[email2b]{Corresponding author, email: sisir.sisirroy@gmail.com.}





\begin{abstract}
We perform new experiment using almost the same sample size considered by Tversky and Shafir to test the validity of classical probability theory in decision making. The results clearly indicate that the disjunction effect depends also on culture and more specifically on gender(females rather than males). We did more statistical analysis rather that putting the actual values done by previous authors. We propose different kind of disjunction effect i.e. strong and weak based on our statistical analysis. 
\end{abstract}


\end{frontmatter}


\section{Introduction}
Tversky and Shafir (1992)~\cite{tversky1992choice} discovered a phenomenon called the {\it disjunction effect}, while following the process of testing a rational axiom of decision theory. It is also called as the sure thing principle (Savage, 1954)~\cite{savage1954foundations}. First consider the states $A$ and $B$ that belong to the state of the world $X$.  This principle states that, if action A over B is preferred, and under the complementary state of the world, again, action A over B is preferred, it is expected that one should prefer action A over B even when the state of the world is not known. Symbolically, this can be expressed as

 If $${P(A\cap X) > P(B \cap X)} $$ and $${P(A\cap X^C) > P(B \cap X^C)}$$

Then $$ P(A) = P(A \cap (X\cup X^C) > P(B \cap (X \cup X^C) = P(B)$$   
i.e. $$ P (A) > P (B) $$ occurs always. 

With the aim of testing this principle, in their experiment, Tversky and Shafir (1992)~\cite{tversky1992choice} performed the test considering a two stage gamble by presenting $98$ students. 
They adopted a two stage gamble, i.e., it is possible to play the gamble twice. The gamble is done under the following two conditions: 
\begin{itemize}
\item The students are informed that they lost the first gamble.
\item The students remained unaware of the outcome of the first gamble.
\end{itemize}

The gamble to be played, had an equal stake, i.e., of wining ${200}$ or loosing ${100}$ for each stage of taking decision, i.e., whether to play or not to play the gamble. 

Interestingly, the results of these experiments can be described into the following manner: 
	\begin{itemize}
		\item The students who won the first gamble 69\% choose to play at the second stage;
		\item The students who lost then 59 \% choose to play again;	
		\item The students who are unaware whether they won or lost 36\% of them (i.e., less of the majority of the students) choose to play again.
			
	\end{itemize}
	
	Explaining the findings in terms of choice based on reasons, Tversky and Shafer (1992)~\cite{tversky1992choice} did raise some questions for these surprising results: whenever, the persons, related to the play, knew that if they win, then as they would have extra house money, they can play again, because if they lose, they can play again to recover the loss. Now the students who did not know the outcome, then the main issue is why sizable fraction of the students want to play again the game since either they win or lose and cannot be anything else? Thus they arrived at the key result, but faced the problem of explaining the outcome just as either a win or loss. Busemeyer, Wang and Townsand (2006)~\cite{busemeyer2006quantum} originally suggested that this finding could be an example of an interference effect, similar to that found in the double slit experiments conducted in modern particle physics. 

Let us consider the following analogy between the disjunction experiment and the classic double slit type of experiment in physics:
Both the cases involve two possible paths: here, in the disjunction experiment, the two paths are inferring the outcome of either a {\it win} or {\it a loss} with the first gamble; for the double split experiment, the two paths are splitting the photon into the upper or lower channel applying a beam splitter. The path taken can be known (observed) or unknown (unobserved), in both the experiments. Finally in both the cases, the fact is that when the case of gambling for disjunction experiment and hence, the detection at a location for the two slit experiments are considered for the chosen unknown, i.e., unobserved conditions, the resultant probability, meant for observing interference phenomena, are found to be much less than each kind of the probabilities which is observed for the known (observed) cases. Under these circumstances, we can speculate that during the disjunction experiment, under the unknown condition, instead of being definite, so far as the win or loss state is concerned, the student enters a superposition state. In fact, this state prevents finding a reason for choosing the gamble. In double-slit experiments, the law of additivity of probabilities of occurring two mutually exclusive events (particle aspect or wave aspect) is violated i.e. total probability
                                                 $$P_{AB} = P_A + P_B$$

for two mutually exclusive events A and B. 
This is due to the existence of interference effects, known as Formula of Total Probability (FTP). It has already been established fact that the two slit (interference) experiment is the basic experiment which violates FTP. Feynman (1951)~\cite{feynman1966feynman}, in many of his works in physics, presented his points with detailed arguments about this experiment. There, the results, i.e., the appearance of interference fringes appeared to him, not at all surprising phenomena. He explained it as follows:
In principle, interaction with slits placed on the screen may produce any possible kind of distribution of points on the registration screen. Now, let us try to explain following quantum probabilistic features which appear only when one considers following three kind of different experiments~\cite{conte2009mental}: 

\begin{itemize}
\item When only the first slit open i.e., the case $B=+1$, in an equivalent manner.
\item When only the second slit is open i.e., $B= -1$, in this case.
\item In the particular case, both slits being open, it is the random variable B determining the slit to pass through.
\end{itemize}

At this stage, let us now choose any point at the registration screen.  Then resultant scenario will be as follows: 
\begin{itemize}
\item the random variable A if  $A= +1$.
\item the opposite case happens if a particle hits the screen at this point, i.e., $A= -1$.
\end{itemize}

Now for classical particles, FTP should predict the probability for the  experiment (both slits are open), supposed to be provided by the (1) and (2) experiments. But, it has already been mentioned that, in case of quantum particles, FTP is violated: for the additional cosine-type term appearing in the right-hand side of FTP, it is the interference effect in probabilities which is responsible. Feynman characterized this particular characteristic feature of the two slit experiment as the most profound violation of laws of classical probability theory. He explained it the following way: 

In an ideal experiment, where there is no presence of any other external uncertain disturbances, the probability of an event, called probability amplitude is the absolute square of the complex quantity. But, when there is possibility of having the event in many possible ways, the probability amplitude is the sum of the probability amplitude considered separately. Following their experimental results, Tversky and Shafir (1992)~\cite{tversky1992choice} demand that this violation of classical probability is also possible to be present, happening in their experiments with cognitive systems. Though, due to the possible restrictions present in quantum mechanics, we could not start from Hilbert formalism at the start, in the laboratory, this formalism was justified by experiments.
Let us now try to interpret the meaning of interference effect within the context of the experiments on gambling, described above. We will follow, here, Busemayer's formulation (2011)~\cite{busemeyer2011quantum} which is as follows:

Two different judgment tasks A and B are considered in this case. The task A is considered having $j$ (taking $j= 2$, binary choice) different levels of response variable and B, with $k$ (say, $k= 7$ points of confidence rating) levels of a response measure. Two groups, being randomly chosen, out of the total participants we have: 

\begin{itemize}
\item Group A gets task A only.
\item Group BA gets task B followed by task A. 
\end{itemize}

The response probabilities can be estimated as follows:
\begin{itemize}
\item From  the group A,  let   $p( A = j)$ be estimated;  This denotes  the probability of choosing level j out of the response to task A .
\item And then from the group B, let $p(B=k)$ be estimated; the corresponding probability denoted  by choosing level k from the task B.
\item Now, it is possible to estimate the conditional probability $p (A=j|B=k)$; which can be stated as the probability of responding with level j from the task A, given the person responded with level k on earlier task B.
\end{itemize}

So, we can write the estimated interference for level j to task A (produced, when responding to task B) as,  
                                     $$ A (j) = p (A = j) â p_T (A = j)$$,
where,  $p_T (A = j)$  denotes the total probability for the response to task A.
In defense of using Quantum formalism in case of human judgments, Busemeyer and Truebold did put four following reasons, beautifully, in their famous paper (Busemeyer et al., 2011~\cite{busemeyer2011quantum}):

{\it (a) judgment is not a simple readout from a preexisting or recorded state, but instead it is constructed from the current context and question; from this first point it then follows that (b) making a judgment changes the context which disturbs the cognitive system; and the second point implies that (c) changes in context produced by the first judgment affects the next judgment producing order effects, so that (d) human judgments do not obey the commutative rule of classic probability theory(Busemeyer et al; 2011~\cite{busemeyer2011quantum})}.
\noindent

In fact, the existence of interference term for microscopic entities or quantum entities clearly indicates the existence of three valued or non-Boolean logic. This is popularly known as Quantum logic. It is mathematically shown that a set of propositions which satisfies the different axiomatic structures for the non-Boolean logic generates Hilbert space structures. The quantum probability associated with this type of Quantum logic can be applied to decision making problems in cognitive domain. It is to be noted that, up till now, no quantum mechanical framework is taken as valid description of the anatomical structures and function of the brain. This framework of quantum probability is very abstract and devoid of any material content. So it can be applied to any branch of knowledge like Biology, Social science etc. Of course, it is necessary to understand the issue of contextualization, for example, here, in case of decision making in brain. It is worth mentioning that the decision making may depend also on culture. For example, the students participated in the above mentioned gamble are mainly taken from the west. So far as the authors' knowledge, concern of this kind of experiment has not been performed taking the subjects from the east. We presume that this is an important factor one should consider in this kind of experiment involving gambling since the very concept of gambling may depend on culture. While experimenting to find the effect of culture, we have found an additional interesting fact. It is not only culture but also gender that might play an important role on disjunction effect. It is natural to expect and consider that the gender dependency is closely dependent on culture. Recently, we performed the above mentioned two stage gambling considering almost the same sample size in India. The results clearly show that the violation of classical probability rule depends very much on the variation of gender. In the next section we will describe the experiment and the results.

\section{Materials and Methods}
 In order to get an insight of the above discussion we have performed a gambling experiment in the line of Tvaersky and Shafir (1952). Our main objective is to see whether there is any disjunction effect, especially in Indian context.

\subsection{Participants}
As we believe that there might be a cultural effect, we carefully select the participants. The participants should consist of both males and females as cultural effect with respect to gender is sometimes significantly observable. We have selected the population of experimental objects as a homogeneous groups with respect to age so that there would not be any effect that can mask our objectives. From Raja Rajeswari Engineering College, Bengaluru, India, we select $101$ college students randomly. It is seen that there are $50$ female and $51$ male students participating the experiment. The students belong 1st, 2nd, or 3rd year of their engineering curriculum and naturally belong to a very homogeneous age group. Prior permission is taken from the principal of the college and the studetns gave consents to this experiment.

\subsection{Design and procedure}
In the experiment we toss a coin and the experiment depends on the outcome. We first divide students into two groups: one consisting $70$ students and the other $31$ students. We perform the experiment similar to that described in Busemeyer. If a students wins the first toss, he/she will receive Rs. $100$; if loses, he/she will lose Rs. $50$. After the first toss, we ask a student whether he/she wants to play again. If he/she wants to play again, he/she will win Rs. $200$ if his/her guess about he outcome matches with the outcome of the second toss; otherwise will lose Rs. $100$. We play this game under two different schemes.

In scheme $1$, each student belonging to the first group ($70$ students), will declare his/her guess about the result of a coin tossing experiment and will be informed the result after the toss. We then ask that student whether he/she wants to play again and record his/her response. In scheme $2$, we perform this experiment for the second group of students ($31$ students) slightly differently than the first scheme. Here, each student will declare his/her guess about the result if the first toss, but he/she will not be informed the result. However, we then ask him/her whether he/she wants to play again and record his/her response. We have used only one coin throughout experiment. Before the beginning of this experiment, we checked whether the coin is unbiased using a binomial experiment and confirmed that it is unbiased. We also make sure that the student who has perfumed the experiment has no way to disclose the result or his/her guess and attitude towards this experimental result. Moreover, there is no exchange of information between the two groups of students for two different schemes.

\begin{itemize}
\item 	We have selected randomly 101 college students from Raja Rajeswari Engineering College, Bengaluru, India. The students belong 1st, 2nd, or 3rd year of their engineering curriculum and naturally belong to a very homogeneous age group.
\item 	We used an unbiased coin for the experiment. Before the experiment we have checked that the coin used is unbiased.
\item We then divide them into two groups: one consisting $70$ students and the other $31$ students.
\item We performed the experiment similar to that described in Busemeyer. If a students wins the first toss, he/she will receive Rs. $100$; if loses, he/she will lose Rs. $50$.
\item Each student belonging to the first group ($70$ students), will declare his/her guess about the result of a coin tossing experiment and will be informed the result after the toss. We record if he/she wants to play again.
\item Each student belonging to the second group ($31$ students) will declare his/her guess about the result if the first toss, but he/she will not be informed the result. We then record if he/she wants to play again.
\end{itemize}
\noindent

{\it {\bf Probability theoretic explanation:}}

We have already discussed that $P(A) > P(B)$ is true always under the conditions that $P(A\cap X) > P(B \cap X)$ and $P(A\cap X^C) > P(B \cap X^C)$. 

Thus, if you prefer A when the event $X$ is known, and if you prefer B when the complementary event $X^C$ is known, then it will imply that you will always prefer A over B irrespective of the events $X$ or $X^C$. Any violation of this is termed as disjunction effect.

We have performed experiment similar to that given in Busemeyer and we observed marked violation of this probabilistic claim and its explanation depends on several factors,  not reported or discussed in literature. Not only the psychological factors, but also other factors like sex, culture, geographical region etc, might have important role to play.

\section{Results}
 We have observed a few interesting facts from the experiment. Instead on looking at the numerical figures of the outcomes of the experiment only as in Tversky and Shafir (1952), we have done a detailed statistical analysis in order to strengthen the interpretation of our observations. As described in the previous section, our experiment consists of two different schemes. Under the first scheme, the students were informed the outcome of the first toss. If they know that they won the first gamble, $76.47\%$ want to play again, i.e. majority want to play again. However we would like to be sure that the result is statistically significant. For this, we perform a test of null hypothesis $H_0:p=0.5$ against an alternative hypothesis $H_1:p>0.5$. The p-value associated with this test is $0.0004$ indicating that the majority wants too play again. On the other hand, if they did not know, whether they won or lost, $58.33\%$ want to play again. Although it seems that majority wants to play again, but the result is not statistically significant (p-value$=0.1215$). Under scheme $2$, when the students did not know the outcome of the first toss, i.e. when they did not know whether they won or lost, $54.84\%$ want to play again. So, in this case, although it seems that majority wants to play again, but the result is not statistically significant since the associated p-value is $0.3601$.

In each case majority wants to play again. But in Busemeyer the corresponding figures are $69\%$, $59\%$, and $36\%$ respectively, while in our experiment the figures are $76.47\%$, $58.33\%$, and $54.84\%$ respectively. The last figure differs widely while the second figure matches surprisingly. We think that probably p-value corresponding to the second figure (i.e. $59\%$) is not significant. So Busemeyer inference needs to be revisited. However, the idea of paying again when they know that whether they won or lost still remains valid if we go only by the actual figures compared to $50\%$, which indicates the state of indifference. So we did an exact test of hypothesis that $H_0: p=0.5$ against $H_1: p>0.5$ i.e. to see absence of disjunction effect. The p-value is $0.3601$ indicating that they are indifferent to the decision. So either disjunction effect is not observed here, or very weak, but they do not prefer to play when they do not know the state.

 It seems clear that the observations by Busemeyer is {\it{not}} matching with ours. We conjecture that this discrepancy may be due to the effect of different cultural settings in which the experiments are conducted. However, we proceed further to examine whether there exists any other factor that might play an implicit role in the experimental results. Indian culture and gender are intermingled always. Hence it would be a wise idea to revisit the experimental results incorporating gender factor. We observed that among females, $75\%$ (p-value$=0.0106$) females want to play again if they know that they won, $52.94\%$ (p-value$=0.3145$) females want to play again if they know that they lost. This result is not at all significant indicating that once lost, females do not want to take one more risk. Moreover, $41.18\%$ (p-value$=0.6855$) females want to play again if they do not the first result.

This picture is markedly different among males. $77.78\%$ (p-value$=0.0038$) males want to play again if they know that they won, $63.16\%$ (p-value$=0.0835$) males want to play again if they know that they lost. However, this result is marginally significant, although not strong. Moreover, $71.43\%$ (p-value$=0.0288$) males want to play again when they do not know the outcome of the first experiment. Since majority wants to play again whether they won or lost or uninformed, there is no disjunction effect observed among males.

\noindent
\vskip5pt

\section{Discussions}
The above results of our experiment raise the following important issues:

\begin{itemize}
\item 	If we consider only absolute values, males do not show any disjunction effect, but females show strong disjunction effect. Now we have to explain this in Indian context, if possible.
\item	However, if we go by the p-values of the corresponding tests of significance, interesting observations can be made using the combination of testsâ results. For all individuals, males and females taken together, let us first combine the results of the conditions that first result is known to be win and that to be loss. The combined p-value for this is 0.00053 indicating that majority of the individuals want to play again if they know the result. But the p-value for the result when they do not know the first result is 0.2366 indicating that there is disjunction effect.
\item For males, disjunction effect is not observed if irrespective of whether we go by the actual values or by the p-values; combined p-value is 0.0029 whereas p-value when the first result is not known is 0.0288. However, the general effect is moving towards the disjunction effect although not established statistically or by observations.
\item For females, although actual observations suggest only weak disjunction effect, but comparing p-values for the combined p-value (0.0223) to the p-value (0.6855) when the first result is unknown shows clear disjunction effect.
\item Disjunction effect is although a clear concept and is realized in a number of experiments, cultural effects are strong especially among males and females. In Indian context, probably disjunction effect for females is so strong that they overcome the absence of disjunction effect among males.
\item Busemeyer's experimental results are based only on the actual values and not tested statistically. This is highly influenced by specific cases and specific scenarios considered in the experiment. It is not wise to declare that the effect is present or absent whenever the number of observations is less or greater than $50\%$; statistical test must be employed to validate and confirm the findings.
\item The missing part of all other previous experiments is the appropriate statistical analysis of the results. Based on our analysis, we propose different kinds of disjunction effect, e.g. strong and weak. 
\end{itemize}
This can be stated in the following manner.

\begin{table}[!h]
\caption{Categorisation of disjunction effect based on actual observation and statistical tests.}\centering
\vspace{2mm}
\begin{tabular}{|c|c|c|}
\hline
\diaghead{\theadfont ColumnmnHead}
{}{Actual observation}  & significant & Not significant\\
{Statistical test}{} & & \\
\hline
Significant & Strong effect & --- \\
(p-value is small) & & \\
\hline
Not significant & Weak effect & No effect \\
(p-value is large) & & \\
\hline
\end{tabular}
\end{table}

The above categorization triggers to rethink the classification of disjunction effect with respect toothed factor like gender etc. Here we propose a general categorization based on actual observations and results of statistical significance tests in presence of another factor `gender'. This is given in Table 2.

\begin{table}[!h]
\caption{A general categorization of disjunction effect based on actual observation and statistical tests.}\centering
\vspace{2mm}
\begin{tabular}{|c|c|c|c|}
\hline

{}{Actual observation} & Significant in both & Significant in males & Not significant \\
\diaghead{\theadfont ColumnmnHead}
{}{} & males and females & but not in females & in both \\
{Statistical test}{} & & or vice versa & \\
\hline
Significant in both & & & \\
males and females & Strong effect & --- & --- \\
(p-value is small) & & & \\
\hline
Significant in males but & & & \\
not in females or vice & Moderate effect & Moderate effect & --- \\
versa (large p-value) & & & \\
\hline
Not significant & & & \\
in both & Weak effect & Weak effect & No effect \\
(p-value is large) & & & \\
\hline
\end{tabular}
\end{table}

In this manner we may come up with more logical categorization.
\noindent
\vskip5pt

\section*{Acknowledgements}
The authors acknowledge the Rector, RR Group of institutions, the principal R.R.Engineering College and the participating students from R.R.Engineering college for their help and cooperation. We would also like to thank Mr. Nepal Banerjee for his help in conducting the experiments. This work is under the project SB/S4/MS:844/2013 approved by SERB, DST, Government of India.



\section*{References}

\bibliography{mybibfile}

\begin{thebibliography}{1}
\expandafter\ifx\csname url\endcsname\relax
  \def\url#1{\texttt{#1}}\fi
\expandafter\ifx\csname urlprefix\endcsname\relax\def\urlprefix{URL }\fi
\expandafter\ifx\csname href\endcsname\relax
  \def\href#1#2{#2} \def\path#1{#1}\fi

\bibitem{tversky1992choice}
A.~Tversky, E.~Shafir, Choice under conflict: The dynamics of deferred
  decision, Psychological science 3~(6) (1992) 358--361.

\bibitem{savage1954foundations}
J.~Savage~Leonard, The foundations of statistics, NY, John Wiley (1954)
  188--190.

\bibitem{busemeyer2006quantum}
J.~R. Busemeyer, Z.~Wang, J.~T. Townsend, Quantum dynamics of human
  decision-making, Journal of Mathematical Psychology 50~(3) (2006) 220--241.

\bibitem{feynman1966feynman}
R.~P. Feynman, R.~B. Leighton, M.~Sands, R.~B. Lindsay, The feynman lectures on
  physics, vol. 3: Quantum mechanics (1966).

\bibitem{conte2009mental}
E.~Conte, A.~Y. Khrennikov, O.~Todarello, A.~Federici, L.~Mendolicchio, J.~P.
  Zbilut, Mental states follow quantum mechanics during perception and
  cognition of ambiguous figures, Open Systems \& Information Dynamics 16~(01)
  (2009) 85--100.

\bibitem{busemeyer2011quantum}
J.~R. Busemeyer, E.~M. Pothos, R.~Franco, J.~S. Trueblood, A quantum
  theoretical explanation for probability judgment errors., Psychological
  review 118~(2) (2011) 193.

\end{thebibliography}

\end{document}